# Magnetic Propeller for Uniform Magnetic Field Levitation

***Mark Krinker***
MKrinker@aol.com
***Alexander Bolonkin***
C&R, 1310 Avenue R, #F-6, Brooklyn, NY 11229, USA
(718) 339-4563, aBolonkin@juno.com, http://Bolonkin.narod.ru

**Abstract**

Three new approaches to generating thrust in uniform magnetic fields are proposed. The first direction is based on employing Lorentz force acting on partial magnetically shielded 8-shaped loop with current in external magnetic field, whereby a net force rather than a torque origins.

Another approach, called a Virtual Wire System, is based on creating a magnetic field having an energetic symmetry (a virtual wire), with further superposition of external field. The external field breaks the symmetry causing origination of a net force. Unlike a wire with current, having radial energetic symmetry, the symmetry of the Virtual Wire System is closer to an axial wire.

The third approach refers to the first two. It is based on creation of developed surface system, comprising the elements of the first two types. The developed surface approach is a way to drastically increase a thrust-to-weight ratio.

The conducted experiments have confirmed feasibility of the proposed approaches.

------------

## Introduction

Overcoming force of gravity was, and remains even today, an ancient human dream. The great portion of energy a human being needs daily is spent for relocation within our Earth's gravity field. Ancient people intuitively understood that overcoming mystical gravity would drastically relieve their survival in any hostile Earthly environment: predators, unfriendly tribes, need to kill animals for nourishment and so on. People envied birds and, understanding that the birds repeal gravity temporarily, somehow, while in the air, people long dreamed about having wings themselves.

This is how concept of levitation settled into human consciousness.

*Levitation: rise or cause to rise despite gravity* (WEBSTER DICTIONARY). Technical progress has partially made this dream a truth. However, since the time of considerable progress in understanding electricity and magnetism in 19th century, humans revealed another media where something (magnetic field) acts on another something (conductor with electric current), causing this the other something to move. Then new dream settled in human consciousness: flying in magnetic field. Unlike air, Earth's magnetic field extends to 1500 km and even more outwards from the planet's surface. Harnessing this force could open the gate for unusual flights. Despite the fact that the force acting on a powered conductor in Earth's magnetic field is insufficient for realizing this dream, works on magneto-levitation already enriched mankind with beneficial achievements.

Known today, magnetic levitation is based on electromagnetic induction in non-uniform magnetic field having a flux density gradient $\nabla\vec{B}$. A time-varying magnetic flux induces eddy current in a conductive object, what results in originating magnetic moment $\vec{m}$ of the object. As a result, a net force $\vec{F} = \vec{m}\cdot\nabla\vec{B}$ origins. The profound review of this technology is made in [1].

However, technical applications of such the method are limited by short-range of the operating non-uniform field as well as the non-uniformity itself.

It is the levitation in weak uniform permanent magnetic fields that holds the great interest today. In particular, levitation in Earth's and cosmic magnetic fields could open the way for a new generation of thrusters and propellers.

However, the induction of Earth's field is pretty insignificant – 30-60 μT, that does not allow to develop needed thrust at reasonable consumption of energy. Moreover, loops with electric current, widely used today for rotational motion in heavy fields, produce a torque rather than a needed net force.

The proposed work shows ways of overcoming such obstacles.

Michael Faraday (1791-1867) was one of the first persons who, almost intuitively, understood that phenomenon of a force, acting on a conductor with electric current in magnetic field, can be explained by distortion of originally symmetric energetic patterns around the wire. Today, we say that the broken symmetry derives free energy and stimulates motion of the active body for minimization of the free energy, - a manifestation of the Le-Shatelier-Braun principle. Origination of well-known forces, acting either on a moving charge or a portion of a conductor with a current is explained by breaking the energetic symmetry. However, due to closed nature of electric current, in a uniform magnetic field a loop with current produces a torque rather than a net force needed for the propelling. Moreover, obtaining needed thrust-to-weight ratio for such the systems are accompanied with generating extremely heavy currents, which is energetically inefficiently. On the other hand, total heavy elemental Ampere's currents in permanent magnets also produce a torque rather than a net force in uniform fields, which block their direct application for obtaining thrust.

In early 70's, the attempt was done to overcome this principal obstacle:

The concept of an apparatus to fly in Earth's magnetic field, comprising rotated dielectric discs with embedded electrets was proposed [11]. As the discs, driven by a turbine, spin in Earth's magnetic field, the Lorentz force, acting on the electrets, origins. The Lorentz force has an opposite direction at opposite ends of the discs. To eliminate the unwanted opposite force, the discs are partially shielded in D-shaped ferromagnetic caves. To compensate an angular momentum, derived by high-speed spinning discs, the discs are rotated in opposite direction. Said system has the following disadvantages:

a) necessity to employ considerable electric charges in electrets, to gain needed Lorentz's force in a weak (30-60 μT) Earth's magnetic field. The total charge needed for that counts for coulombs, that produce forces of electrostatic repulsion which can break the discs;
b) mechanical instability of the system, rotating heavy discs at high angular speed;
c) gradual discharge of the electrets by ionizing radiation;
d) limited altitude achieved by the apparatus over Earth, because of inability a gas turbine to operate in a thin air.

The proposed methods and apparatuses eliminate shown drawbacks and make magnetic propelling feasible.

## 1. Current Contour Having Magnetically Shielded Portion in External Magnetic Field

The method shown in Fig.1 is based on the following: close electric contour 10 with current, driven by the source 11, has its portion inside a hollow ferromagnetic body 12, [2]. The external magnetic field 13 acts only on the external portion of the contour 10 – there is actually no field inside the hollow ferromagnetic body 12. As the results, the Lorentz force 14 origins. The force acts only on the exposed portion of the contour 10. The value of the Lorentz force is $\vec{F} = I\left[\vec{l} \times \vec{B}\right]$, where $I$ – current, $l$ –length of exposed portion of the conductor 10. In particularly, if the external magnetic field is formed by Earth, having induction $B$ = 30-60μT, then each 1m of the exposed conductor with 1A of the flowing current experiences action of $(3 \div 6) \cdot 10^{-5} N$, depending on the longitude.

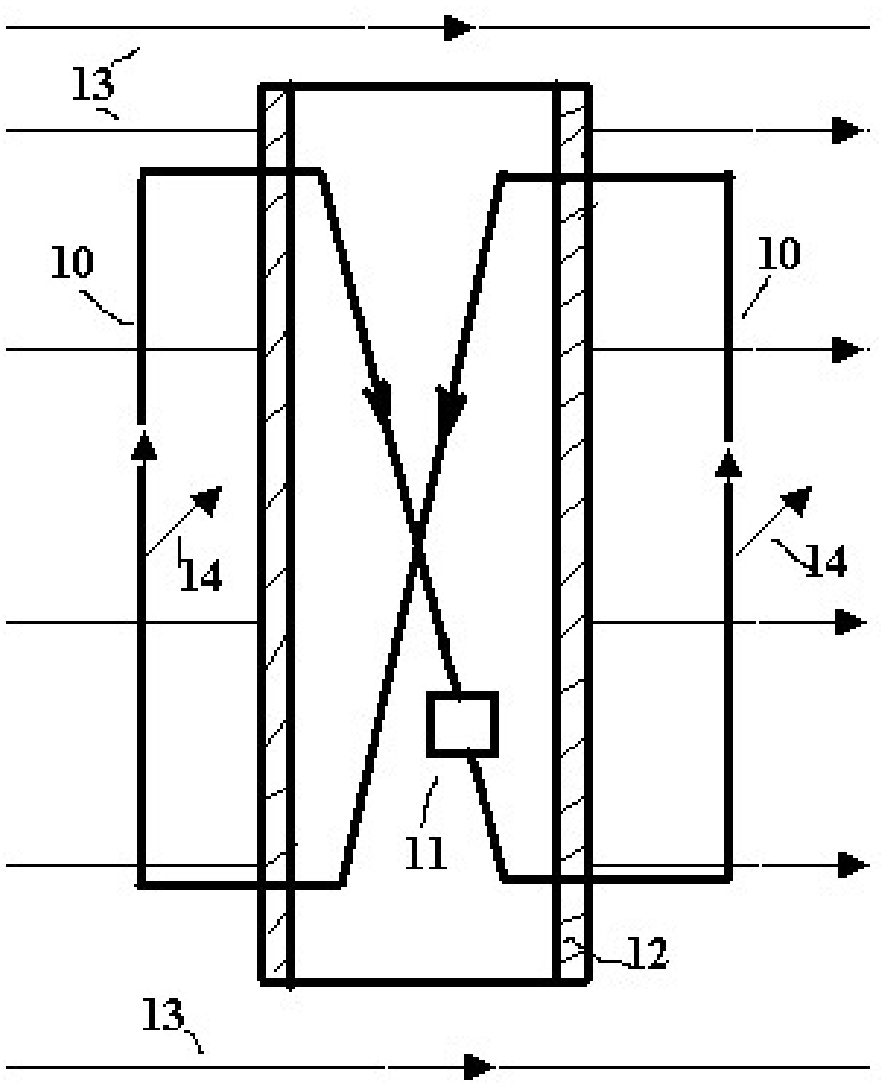

**Fig.1**

General concept of generating a net force acting on a partially shielded closed contour with current in external magnetic field. *Notations:* 10 - twisted 8-shaped conductor with current; 11- source of current; 12- hollow tube-like ferromagnetic body; 13 - external magnetic field; 14 - net Lorentz force.

To increase the thrust, the whole contour 10, or at least its portion, which is exposed to magnetic field, can be made of a mu-metal, having considerable magnetic permeability. The mu-metal wire produces additional magnetic field around it and, therefore, enforces the thrust.

**1.1. Developed Surface 8-Shaped System**

Taking into consideration that thrust is developed by 2 field-exposed conductors 10, each having length $l$ and weight $Pc$, while the total weight depends on 4 involved conductors, the efficiency of the system shown above is $\frac{2IlB}{4P_c + P_t} = \frac{2IlB}{P}$, where $Pt$ is a weight of the shielding tube 12.

Value of the total weight $P = 4\pi r^2 l\rho_1 g + 2\pi r_1 ld\rho_2 g = 2\pi gl\left(2r^2\rho_1 + r_1 d\rho_2\right)$. The efficiency equals

$$Eff = \frac{IB}{\pi g\left(2r^2\rho_1 + r_1 d\rho_2\right)}, \qquad (3)$$

where $g, r, r_1, \rho_1, \rho_2, d$ are free-fall acceleration, radius of wire 10, radius of tube 12, density of wire 10, density of tube 12, and thickness of the tube 12, respectively.

Considering the minimal radius of the tube $r_1 = 2r$ and its thickness $d = r$, the efficiency can be shown as

$$\frac{IB}{\pi g\left(2r^2\rho_1 + r_1 d\rho_2\right)} = \frac{IB}{2\pi r^2 g\left(\rho_1 + \rho_2\right)} \qquad (4)$$

From here, for this system to be efficient, the condition $I_{cr} \geq 2\pi r^2 g\left(\rho_1 + \rho_2\right)B^{-1}$ has to be observed. Having $\rho_1 = 8.93 \cdot 10^3 kg/m^3$ (copper), $\rho_2 = 7{,}88 \cdot 10^3 kg/m^3$ (iron), $B = 5 \cdot 10^{-5} T$ (Earth), and the radius of the wire 10, $r = 10^{-5} m$, the critical current is 2.1A. The less diameter of the wire 10, the less is critical current.

The phenomenon is explained due to the fact that the acting force is proportional to the surface on which it acts, while the weight is proportional to the volume. The smaller the radius, the more surface-to-volume ratio resulting. This is a developed surface system. The apparatus of Fig.2 can be made of arrays of the developed surface modules, like the one calculated above, fed by a common power bus. Beside, the contour 10 and the tube 12 can be made of a μ-metal to increase the magnetic field of the contour 10 as well as efficiently shielding the external field 13.

### 1.2. Earth's Magnetic Field and Lighting

A role of Earth's magnetic field usually is ignored in understanding of a lighting discharge. For a forked lighting, the discharge consists of a leader stroke, building steadily downward, and a return stroke, which runs from ground to clouds. The leader stroke progresses toward the ground by approximately 100m steps, lasting 2 us with 50 us pause. The discharge current can reach 20000 A.

We believe that division the lighting cord for the steps is caused by Earth's magnetic field. Indeed, at Earth's magnetic induction as much as 50 μT, each the meter of the lighting cord experiences action of 1N force, acting mostly on the ions. This force, acting normally to the cord, simply breaks it. Then the discharge stops and then electric field strength raises until a new breakdown and the following discharge. Then the process repeats consequently. A front of the stroke moves at 15000 m/s, thus the ions experience an acceleration as much as $10^8 m/s^2$ between collisions. The rate of the acceleration is enough to rapidly break the stroke.

### 1.3. General Concept of the Apparatus Having Partially Shielded Current Contour in External Magnetic Field.

Fig. 2 shows a general concept of the aircraft employing the contour, having the magnetically shielded portion. The apparatus to fly in external magnetic field comprises a vertical thrust platform 20 with the exposed to the field portions of the 8-shaped contour with currents 21, the horizontal thrust platform 22 with the exposed portions of the 8-shaped contour with currents 23, and magnetic stabilizers 24 to align the aircraft for positioning the contours 21 and 22 normally to magnetic lines of force of the field 25.

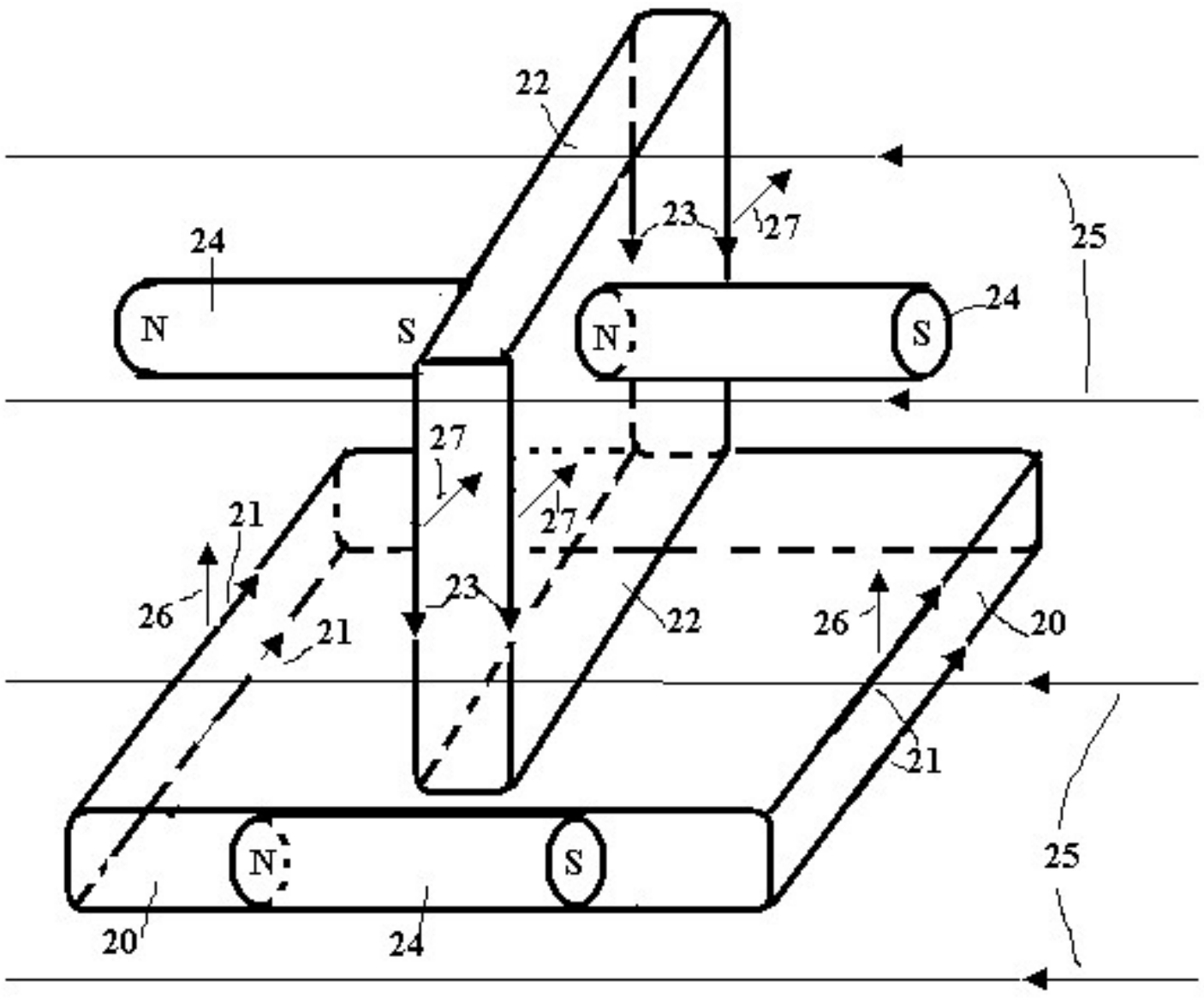

**Fig.2.**

General concept of a thruster employing current contours having magnetically shielded portions in external magnetic field. *Notations:*

20-horizontal platform comprising 8-shaped powered conductors having magnetically-shielded internal portions; 21- exposed to external magnetic field portions of the 8-shaped horizontal conductors; 22- vertical platform comprising 8-shaped powered conductors having magnetically-shielded internal portions; 23- exposed to external magnetic field portions of the 8-shaped vertical conductors;

24- magnetic stabilizers, having magnetic moment, which is anti-collinear to the external magnetic field; 25- external magnetic field; 26- vertical net force acting on the field-exposed conductors 21 of the horizontal platform 20; 27- horizontal net force acting on the field-exposed conductors 23 of the vertical platform 22

As the result, two Lorentz forces origin: force 26 acts on the contour 21 vertically, while the force 27 acts on the contour 22 horizontally. Therefore, both vertical and horizontal thrusts origin. The sources of the currents 21 and 23 are located inside the platforms 20 and 22 (the sources are not shown here). The currents 21 and 23 form 8-shaped contour of the Fig. 1. Therefore, the total shape of the currents 21 and 23 does not correspond to shape of the platforms 20 and 22.

The stabilizers 24 are heavy field magnet, having lines of force aligned along North-South direction. Actually, it conducts itself like a needle of a giant compass, always keeping position of the apparatus stable in Earth's field.

**1.4. Magnetic Shielding. Materials for Shielding Device.**

Operation of the 8-shaped system of the Fig.1 depends on shielding of the portion of the current contour 10. First of all, it has to be said that unlike electrostatic shielding there is no ultimate magnetic shielding. This fundamental difference is caused by a different physical nature of conservative electrostatic field and solenoid magnetic field. There are no "magnetic charges" at which magnetic lines of force start and end like it has place when shielding conservative electric fields by conductive materials having free electric charges.

The shielding action of magnetic screens is based on refraction of magnetic lines of force due to magnetization of the material of the shield in external magnetic field. The magnetic field developed by the shield opposites the original external field in vicinity of the walls of the shield. Due to closed nature of magnetic lines of force, their superposition in external portion of the shield results in that the strength of the resulting field is reduced in immediate vicinity of the shield while the field inside the walls of the shield is elevated. It looks like refraction of the magnetic lines of force.

Successful screening external magnetic field implies immediate reaction of the magnetic shield to external field strength: the magnetic inductance of the material of the shield has to follow changes of the external field.

Because the induction of the field inside the material is directly proportional to relative magnetic permeability $\mu_r$, then the capability to magnetic shielding depends on this property of the material:

$$\vec{B} = \mu_0 \vec{H} + \vec{M} = \mu_0 \mu_r \vec{B},$$

where $M$ is a magnetization of the material of the shield.

However, magnetization experiences saturation in heavy external fields.

Due to that, the permeability is a sub-linear function of the magnetic field strength.

From this, it follows, that saturated magnetic shield can't perform its function.

Selecting materials for magnetic shielding depends on the fact that a relative magnetic permeability depends on a flux density, the induction, and changes with it. For this reason, relative permeability also changes: the raise of the temperature usually decreases the relative permeability. For instance, if the temperature increases from 20 to 80 centigrade, then a typical ferrite can suffer a 25% permeability drop.

Fig.3 shows permeability and inductance versus field strength for iron. As seen from the figure, magnetic permeability is maximal at intensive increase of inductance vs. field strength. This is why the relative permeability can be evaluated as a derivative: $\mu_r = \frac{dB}{dH}$.

Magnetic permeability also depends on a previous mechanical working, which defines a microstructure. It's especially considerable for transformer steel. The mechanical working orients grains along one direction (anisotropy) giving increased permeability.

The table bellow shows both dimensional magnetic permeability $\mu(H \cdot m^{-1})$, (Henrys per meter) and relative magnetic permeability $\mu_r = \mu / \mu_0$.

Approximate maximum permeabilities

| Material | $\mu$/(H m$^{-1}$) | $\mu_r$ | Application |
|---|---|---|---|
| Ferrite U 60 | 1.00E-05 | 8 | UHF chokes |
| Ferrite M33 | 9.42E-04 | 750 | Resonant circuit RM cores |
| Nickel (99% pure) | 7.54E-04 | 600 | - |
| Ferrite N41 | 3.77E-03 | 3000 | Power circuits |
| Iron (99.8% pure) | 6.28E-03 | 5000 | - |
| Ferrite T38 | 1.26E-02 | 10000 | Broadband transformers |
| Silicon GO steel | 5.03E-02 | 40000 | Dynamos, mains transformers |
| supermalloy | 1.26 | 1000000 | Recording heads |

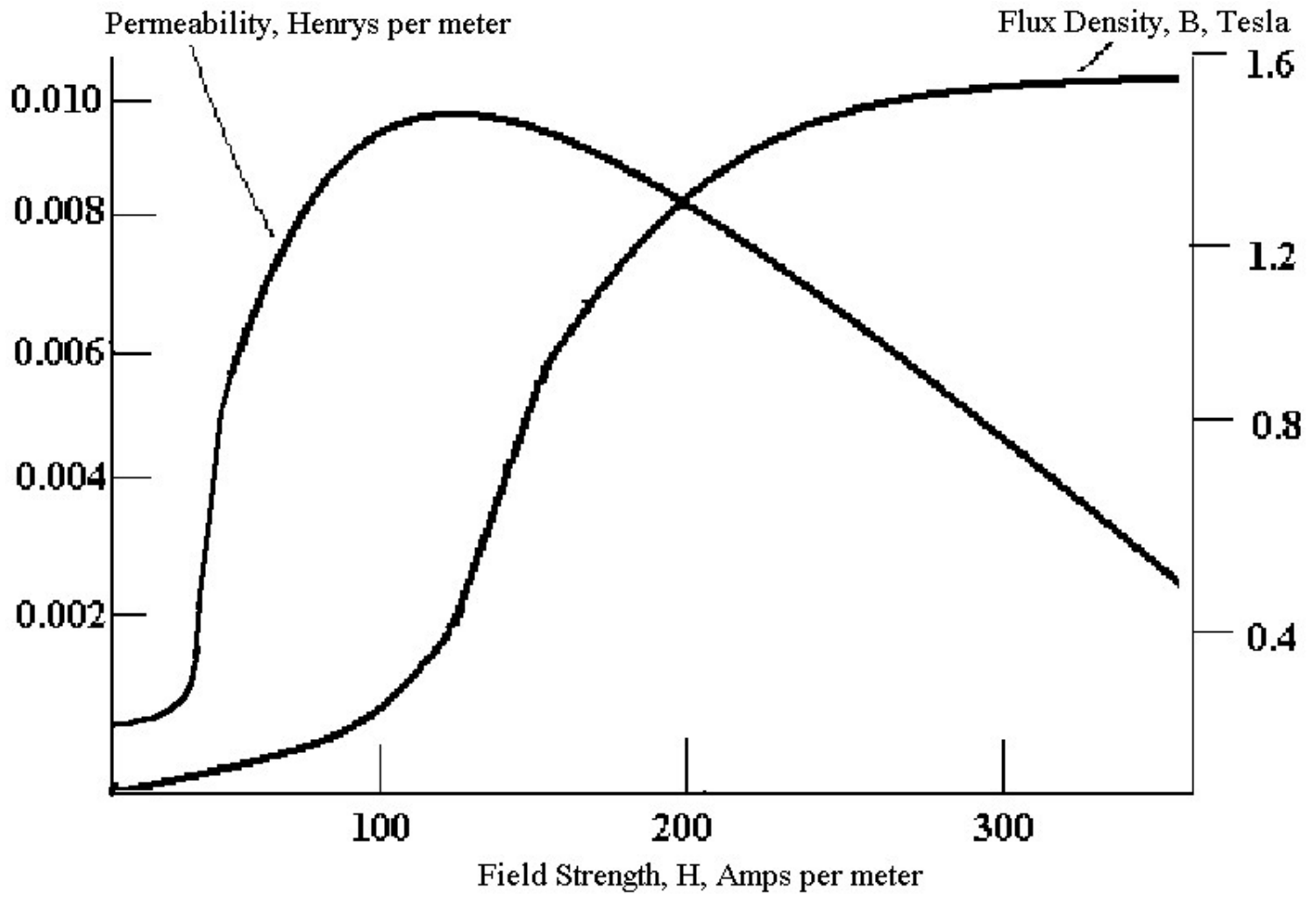

**Fig.3**
Permeability and induction vs. field strength for iron.

Selection materials for magnetic shielding have to take into consideration operating field strength developed by the current contour and visa versa. Therefore, the field strength developed by a shielded portion of the contour 10, Fig.1 has not to exceed some limit, depending on the material. Otherwise, the shield 12 becomes transparent for external magnetic field and the net force 14 experiences reduction. However, too weak field, as this follows from the Fig, is also unacceptable because reduced magnetic permeability.

Operating field, developed by conductors 10 of 8-shaped system counts tens-hundreds of A/m. Based on the table shown directly above, it follows that the best materials for the shielding are those having industrial power rates: ferrite N41, iron, ferrite T38, silicon steel. Supermalloy simply will lose its splendid shielding properties in such fields. However, this does not deny possibility of employing this material in the lower field of the Developed Surface 8-Shaped System.

## 2. Virtual Wire System: Employing the Own Magnetic Field of an Object

Another method to develop lift and moving forces in magnetic field is the developing of the difference of density of magnetic energy in vicinity of an object by means of employing magnetic field of the object rather than employing powered wire, Fig.4, [3,12]. The systems, realizing this approach can be subdivided for symmetric and asymmetric ones in terms of its physical structures.

### 2.1. Symmetric Virtual Wire Systems.

Breaking energetic symmetry is a cause of originating forces.

Force *F*, acting on a border of two bodies, having different densities of energy $w_1$ and $w_2$, is $F_n = \iint_S \Delta w(x,y)dS$ where *S* is a total area of surface of the adjacent areas of energy.

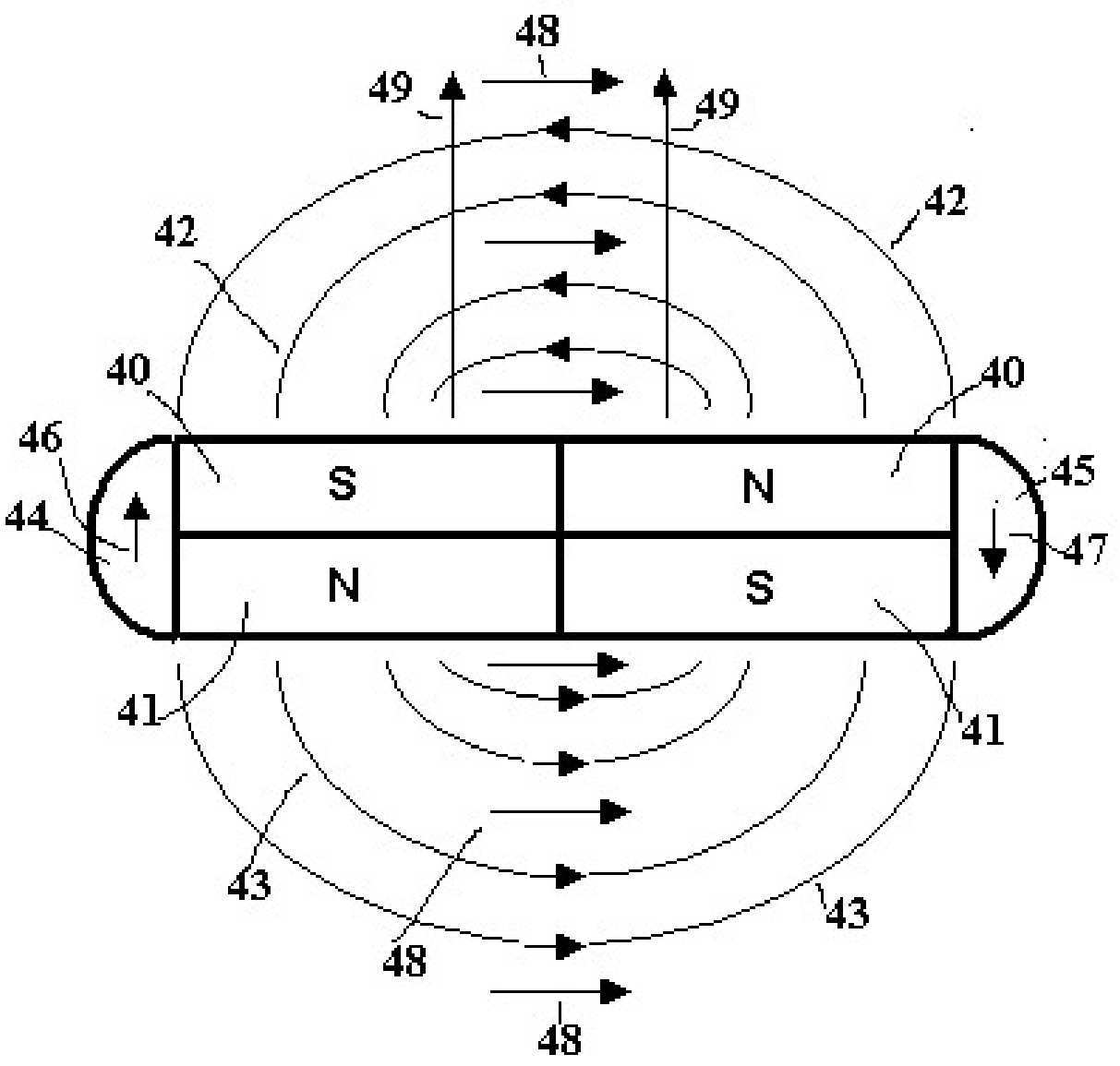

**Fig. 4a.**

Physical base of the virtual wire system: *Notations*: 40 - bar magnet; 41 - bar magnet, having magnetic moment, which is anti-collinear to the magnetic moment of the magnet 40; 42 - magnetic field of the magnet 40; 43 - magnetic field of the magnet 41; 44 - left magnetic shunt for "absorbing" end magnetic field of the magnets 40-41; 45 - right magnetic shunt for "absorbing" end magnetic field of the magnets 40-41; 46 - left-end magnetic field inside the shunt *(direction in Fig.)*; 47 - right-end magnetic field inside the shunt; 48 - external magnetic field; 49 - net force.

The magnetic assembly of the magnets 40-41 produces opposite-directed fields 42-43, which are symmetric in terms of energy while no external filed is applied. Superposition of the external field 48 breaks the energetic symmetry, developing the net force 49.

Origination of this force is based on the Le Chatelier-Braun principle: When a constraint is applied to a dynamic system in equilibrium, a change takes place within the system, opposing the constraint and tending to restore equilibrium. As far as the considered system is concerned, the equilibrium implies equality of density of magnetic energy around the object.

Fig. 4a shows a system, comprising two magnets 40 and 41, arranged in anti-parallel way, producing magnetic fields 42 and 43, respectively. The fields 42 and 43 have equal induction BM. To eliminate penetrating unwanted opposite magnetic fields from the ends of the magnets 40 and 41 into external media, magnetic shunts 44 and 45 are employed. The fields 46 and 47 are canalized inside the shunts 44 and 45, which made of ferromagnetic materials. The external magnetic field 48, having induction BE, is subtracted from the filed 42 and is added to the field 43.

Fig.4b illustrates breaking initial axial energetic symmetry in vicinity of the Virtual Wire System as the external field **Be** is applied.

As seen from the Figs. 4a and 4b, the intrinsic vector field **Bm** of the magnetic assembly 40-41and the external field **Be** may be added or subtracted.

Let's consider energetic conditions of the upper and lower planes of the Fig. 4 taking into consideration that an angle α between **Bm** and **Be** is a function of a coordinate on *x-y* plane. Taking the angle α of the lower plane, having parallel projection of Bm and Be, as a common variable for both the planes, we have:

the resulting field under the lower plane, $B_{r+} = \sqrt{B_m^2 + B_e^2 + 2B_m B_e \cos\alpha_{x,y}}$ , and the resulting field over the upper plane, $B_{r-} = \sqrt{B_m^2 + B_e^2 - 2B_m B_e \cos\alpha_{x,y}}$ .

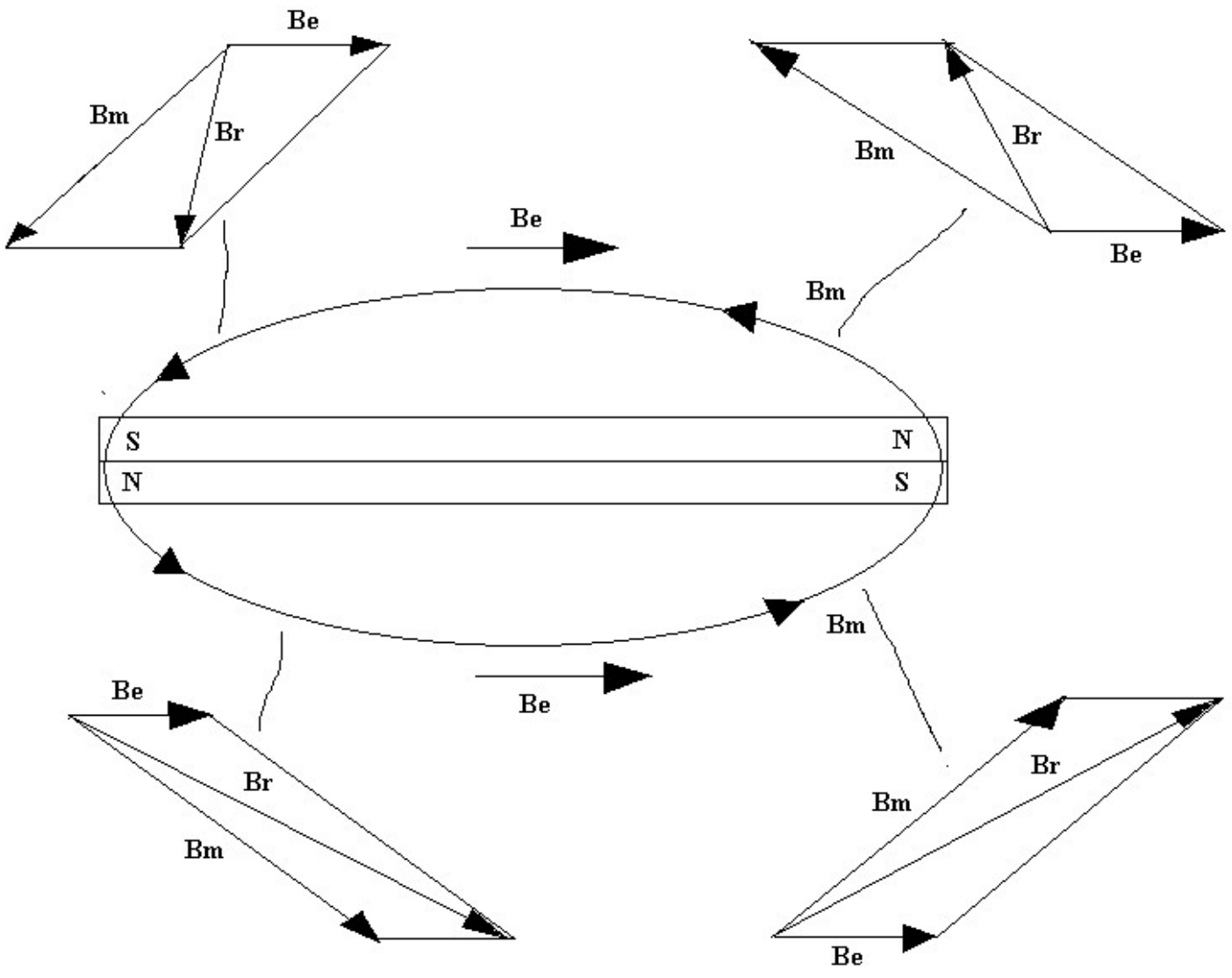

**Fig.4b**

Breaking initial axial energetic symmetry of the virtual wire system as the external field **Be** is applied. The resultant vector **Br** is a result of superposition of the vectors **Be** and **Bm .**

The gained excessive density of magnetic energy under the lower plane is

$$w_+ = \frac{B_{r+}^2}{2\mu_0\mu_r} = \frac{B_m^2 + B_e^2 + 2B_m B_e \cos\alpha_{x,y}}{2\mu_0\mu_r}$$

The depleted density of magnetic energy over the upper plane is

$$w_- = \frac{B_{r-}^2}{2\mu_0\mu_r} = \frac{B_m^2 + B_e^2 - 2B_m B_e \cos\alpha_{x,y}}{2\mu_0\mu_r}$$

The difference of the densities, taken normally to *x-y* plane along *Z*-axis is

$$\Delta w = w_+ - w_- = \frac{2B_m B_e \cos\alpha_{x,y}}{\mu_0\mu_r}$$

The value of $\Delta w$ is a density of free energy of the system. It derives a net force *F* directed into area of the depleted energy.

Generally, the distribution of the free energy depends on angle α between the external field 48 and intrinsic fields 42 and 43 of the magnet:

$$\Delta w(\alpha) = \frac{2}{\mu_0\mu_r} B_E B_M \left|\cos\alpha\right| \qquad (5)$$

The difference of densities of energy *Δw* reaches maximum at the center of the system. Said the difference of densities of magnetic energy in vicinity of the magnets 40 and 41 results in origination of normally-acting force 49, which can be calculated as

$$F_n = \iint_S \Delta w(x,y)\, dS = F = \frac{kB_E B_M S}{\mu_0\mu_r}, \qquad (6)$$

where $1 \le k \le 2$, depending on proportion of the object.

**2.2. Equilibrium of the Virtual Wire System in Magnetic Field.**

Beside the produced net force, virtual wire systems have own magnetic moments, which develop torques in external magnetic fields. This affects the equilibrium of the magnetic propeller and has to be taken into consideration.

Any rotation in external magnetic field is accompanied by variation of projection of the vector of the external field onto direction of the internal field of the magnets of Fig.4. As this takes place, acting value of the external vector is $\vec{H} = \vec{H}_0 \cos\alpha$, where α is an angle between magnetization and external field during rotation. Energetically, the process of the rotation is asymmetric due to hysteresis, Fig.6. The area of hysteresis loop is an additional energy, which contributes into general energetic asymmetry of the virtual wire defining the equilibrium angle.

### 2.2.1. Equilibrium in a Vertical Plane in Gravity Field

According to Le Chatelier-Braun principle, the system of Fig. 4 can restore the initial equilibrium, in which it rested before superposition of the field 48, by a linear motion or self-turning a system until the densities of magnetic energies get equal.

Fig. 5 shows the system of Fig. 4, having horizontal axis or rotation in a gravity field. The magnetic system 50, having its own fields 51, $B_e$, placed in external field 52, $B_m$. The free energy of the system originates in the volume of the magnets 50. The free energy of the system 50 forces it to turn around an axis 53. Then, its original center of mass 54 shifts to new position 55. The center of mass is shifted from the axis 53 for a distance 56, *h*. The value of its vertical shift is $\Delta h$. Then, the potential energy increases for $\Delta W = mg\Delta h$, which depends on the equilibrium angle $\beta$. The equilibrium takes place at the equal free energy and additional potential energy. From here

$$\frac{kB_E B_M S}{\mu_0 \mu_r}\cos\beta = F\cos\beta = 2mg\sin^2\left(\frac{\beta}{2}\right) \quad . \qquad (7)$$

From here, the originated magnetic force can be calculated as

$$F = \frac{2mg\sin^2\left(\frac{\beta}{2}\right)}{\cos\beta} \quad . \qquad (8)$$

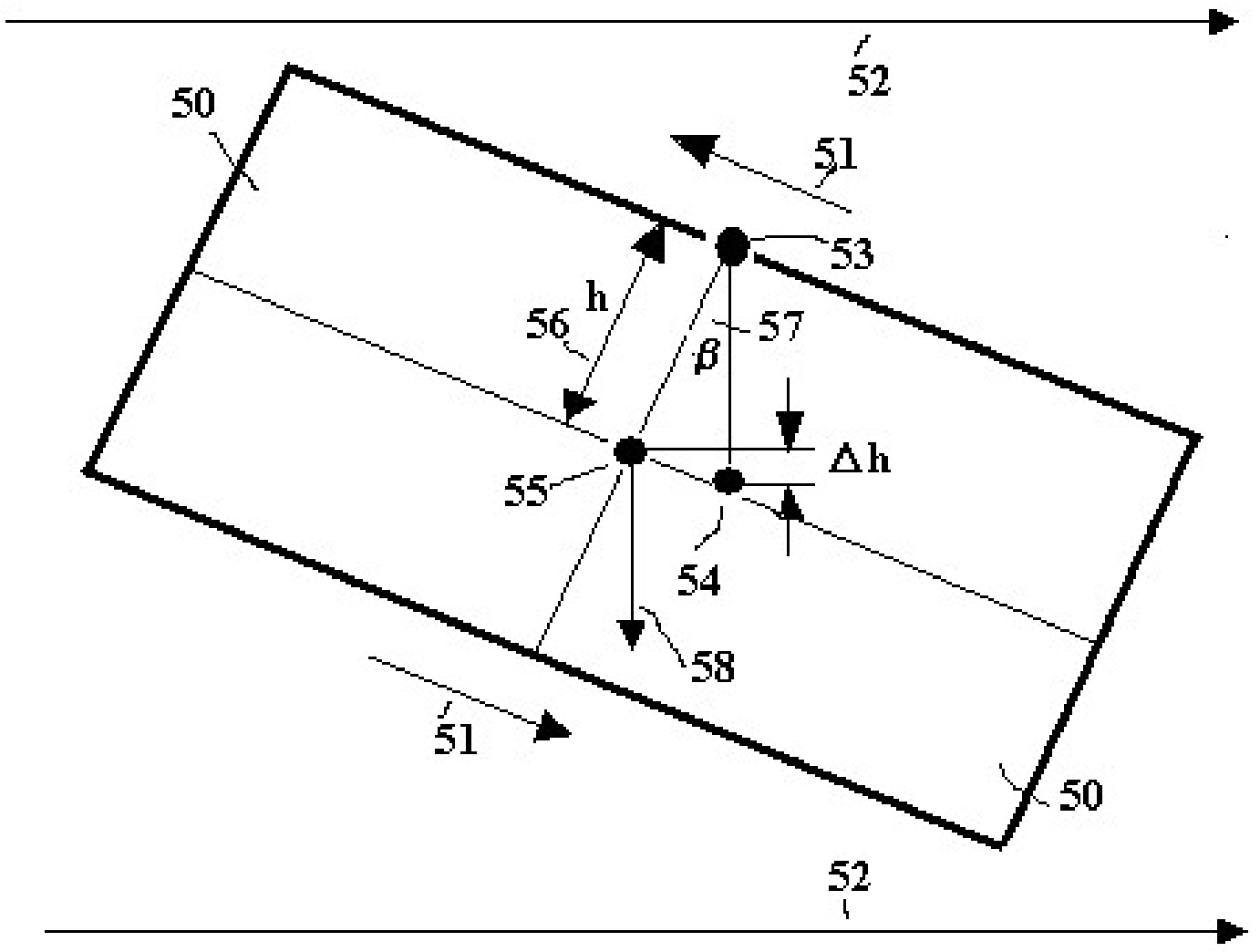

**Fig. 5.**

Equilibrium of the virtual wire system in gravity field (vertical plane). *Notations*: 50 - assembly of anti-collinear bar magnets; 51 - intrinsic magnetic fields of the assembly; 52 - external magnetic field; 53 - axis of rotation; 54 - position of center of mass before rotation; 55 - position of the center of mass after rotation-equilibrium position; 56 - distance between the center of mass and the axis of rotation; 57 - angle of rotation – the equilibrium angle; 57 - gravity force applied to the center of mass.

Asymmetry of energy caused by hysteresis processes taking place during rotation was ignored here. However, it's justified here taking into consideration that a gain of gravitational energy exceed that of the hysteresis process. However, the hysteresis is a major factor as it comes to rotation in a horizontal plane, where no shift of the center of mass takes place,

### 2.2.2. Rotation around Vertical Axis

Application of external magnetic field *H* causes rotating anti-parallel magnetic duplex (virtual wire system) in a horizontal plane until it reaches some equilibrium angle $\alpha$ between direction of magnetization $M$ of the magnets 40-41 and the external field $H$. The process is initiated by a field-induced braking symmetry of magnetic moments. In reality, magnetic moments of the magnets 40-41 are equal only until the external field is applied. The following energetic situation can be illustrated by Fig. 6.

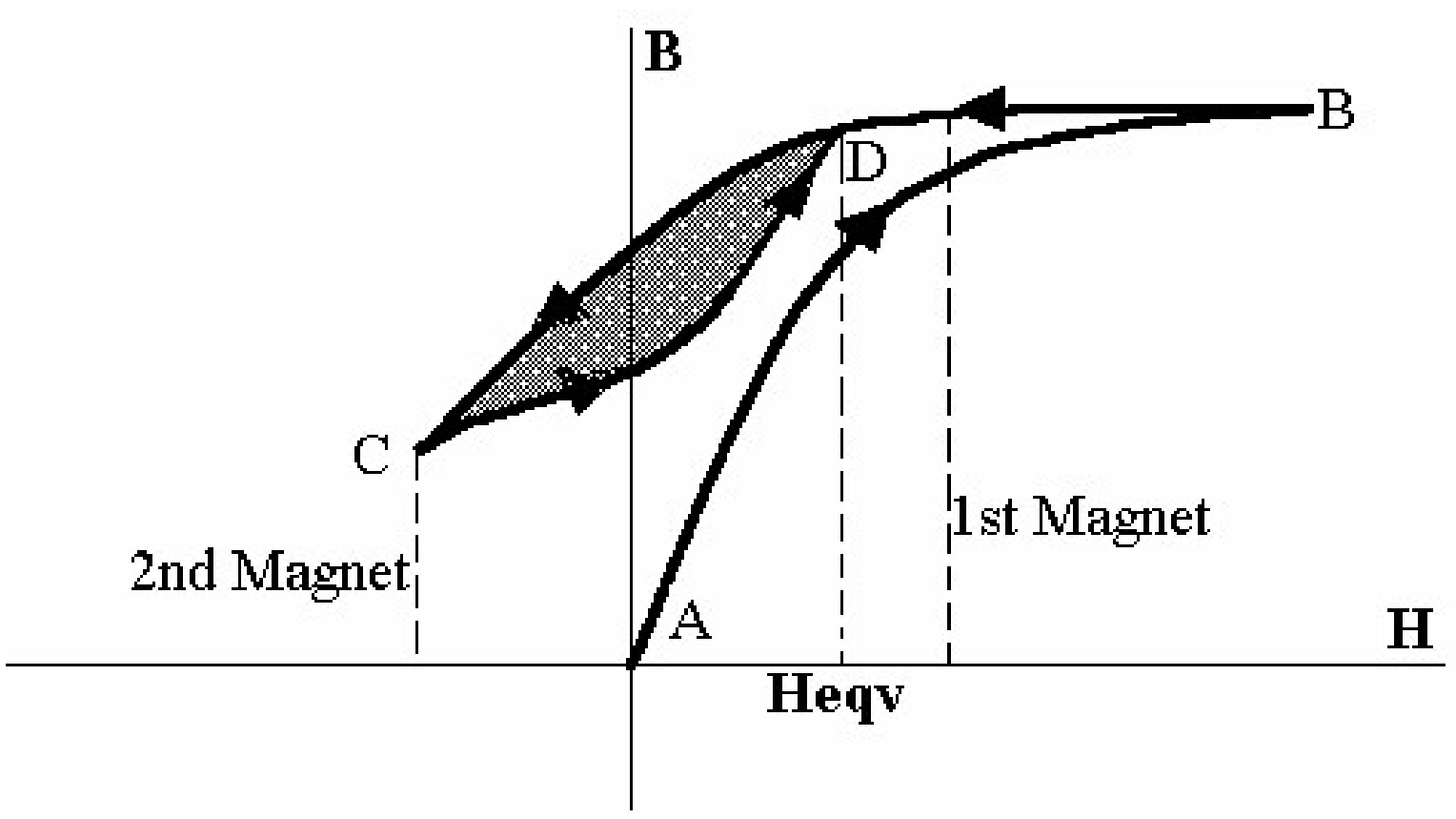

**Fig.6.**
Influence of hysteresis of the magnetic assembly on equilibrium of the virtual wire system.

The first magnet (41 of Fig.4) runs way BD during rotation, while the second magnet (40 of Fig.4) runs the lower branch of CD. Due to the hysteresis, their inductions meet each other at some point D, which corresponds to the equilibrium field strength $\vec{H}_{eqv} = \vec{H} \cdot \cos\alpha_{eqv}$. The diagram explains why the equilibrium angle between the field and the magnetic moments can't be 90 degrees ($H = 0$): the inductions are not equal at this angle.

The first magnet, 41, is in energetically unprofitable condition: it has a positive potential energy $U = -\vec{m} \cdot \vec{B}$ in the external field, while the second magnet, 40, has a negative potential energy, resting in energetically profitable condition before combining both the magnets into a whole system. As it follows from Fig.4 magnetic moment $m$ of the first magnet exceeds that of the second magnet that causes the system to start turning for minimizing its free energy. As placing in the external field $H$, the second magnet runs a path BC on the Fig. 6. As rotating in the external field, the second magnet 40 moves from the point C in B-H plane, while the first magnet, 41, experiences the path BD. Due to nonlinear properties of a hysteresis, the opposite paths of the both magnets do not co- inside until some point is reached D, where their inductions become equal. At this point *Heqv* their energies get equal and the system stops the rotation. While rotating, the magnetic duplex performs work $W = \int_C^D B_1 H dH - \int_D^C B_2 H dH$ within contour CD, where $B_1$ and $B_2$ are hysteresis branches for the magnets 1 and 2 respectively. The field $H = H_{\max} \cos\alpha$ and, therefore, the equilibrium angle $\alpha$ depends exquisitely on magnetic characteristics of the employed magnets. From the specific properties of the hysteresis, it follows that $\alpha$ never can be neither 90 nor 0 degrees.

**2.2.3. Developed Surface Virtual Wire System.**

Force, produced in the virtual wire system depends on a total border area between the object and the media of its motion. The less physical size $R$ of the object, the more surface area-to-volume ratio, $1/R$. On the other side, weight is a function of a volume. Therefore, to make a thrust-to-weight ratio more efficient, surface-to-volume ratio has to be as maximal as possible.

Development of this approach is shown in Fig.7. This system comprises an array of oblong film-like permanent magnets arranged in anti-parallel way. Their $l >> p$. Because of the total energy of end area drastically less than that of major portion of the film magnets, there is no need in magnetic shunts for elimination of end field in surrounding media. The total area of surface of a stratified system is $S = Nns$, where $N$-is a number of layers, $n$ is a number of magnetic pairs in one layer, and $s$ is an area of one magnetic pair. For one anti-parallel magnetic pair, thrust-to-weight ratio can be shown as

$$\frac{F}{W} = \frac{2B_M H_E S}{V\rho g} = \frac{2B_M H_E lw}{lwd\rho g} = \frac{2B_M H_E}{d\rho g}, \qquad (9)$$

where $l$, $w$, $d$, $\rho$ and $g$ are a length, width, thickness, density and acceleration of free fall, respectively. For this system being feasible, $F/W$ has to exceed 1. Therefore, at permanent $B$ and $H$, critical thickness $d_{cr} = \frac{2B_M H_E}{\rho g}$. From here, for iron–film ($\rho = 7.8 \cdot 10^3 kg/m^3$) magnets, having induction in immediate vicinity of the surface, $B$ = 0.1 T and Earth's field strength 30 A/m, $d_{cr} = 7.7 \cdot 10^{-5} m = 77 \mu m$, that is, if $d$ is less than this value, the system can float in Earth's magnetic field. However, the calculation implies that length $l >> w,d$, otherwise magnetic shunts has to be employed to eliminate unwanted side field.

The system's payload, sometimes called "carrying capacity", can be found as

$$P = F - W = 2NnsB_M H_E - Nns\rho d g = Nns(2B_M H_E - \rho d g). \qquad (10)$$

Here $N$ and $n$ are numbers of the layers and number of magnetic duplets in each the layer, while s is an area of each the duplet. For the system, having $N$ = 1000 and $n$ = 1000, with $s = 1m \cdot 10^{-3} m$, $d = 3 \cdot 10^{-5} m$, and the rest of parameters like shown above, $P = 3.66 \cdot 10^3 N = 370\ kg$. Area of one layer is $1m^2$. The layers are arranges one under another. The following fact has to be taken into consideration:

The $1m^2$ **base** comprises batch-arranged $N = 1000$ layers, while each of the layer counts $1m^2$. So, the total surface "hidden" over 1 $m^2$ - base is $10^3 m^2$. <u>This explains the seeming paradox of the abnormal thrust: it's actually calculated for $10^3 m^2$ effective area.</u>

So, one square meter of surface of the Developed Surface Magnetic System, having a vertical batch of 1000 layers, can lift 370 kg of payload in Earth's magnetic field if the induction developed by film-like magnets 60-61 is 0.1T.

The value of this induction is not abnormal for rare Earth magnet. However, the film-like magnets have to be mono-crystal rather than amorphous structures to develop this induction. The total thickness of 1000 of $10^{-4} m$ layers (including thickness of a plastic carrier) is 10 cm.

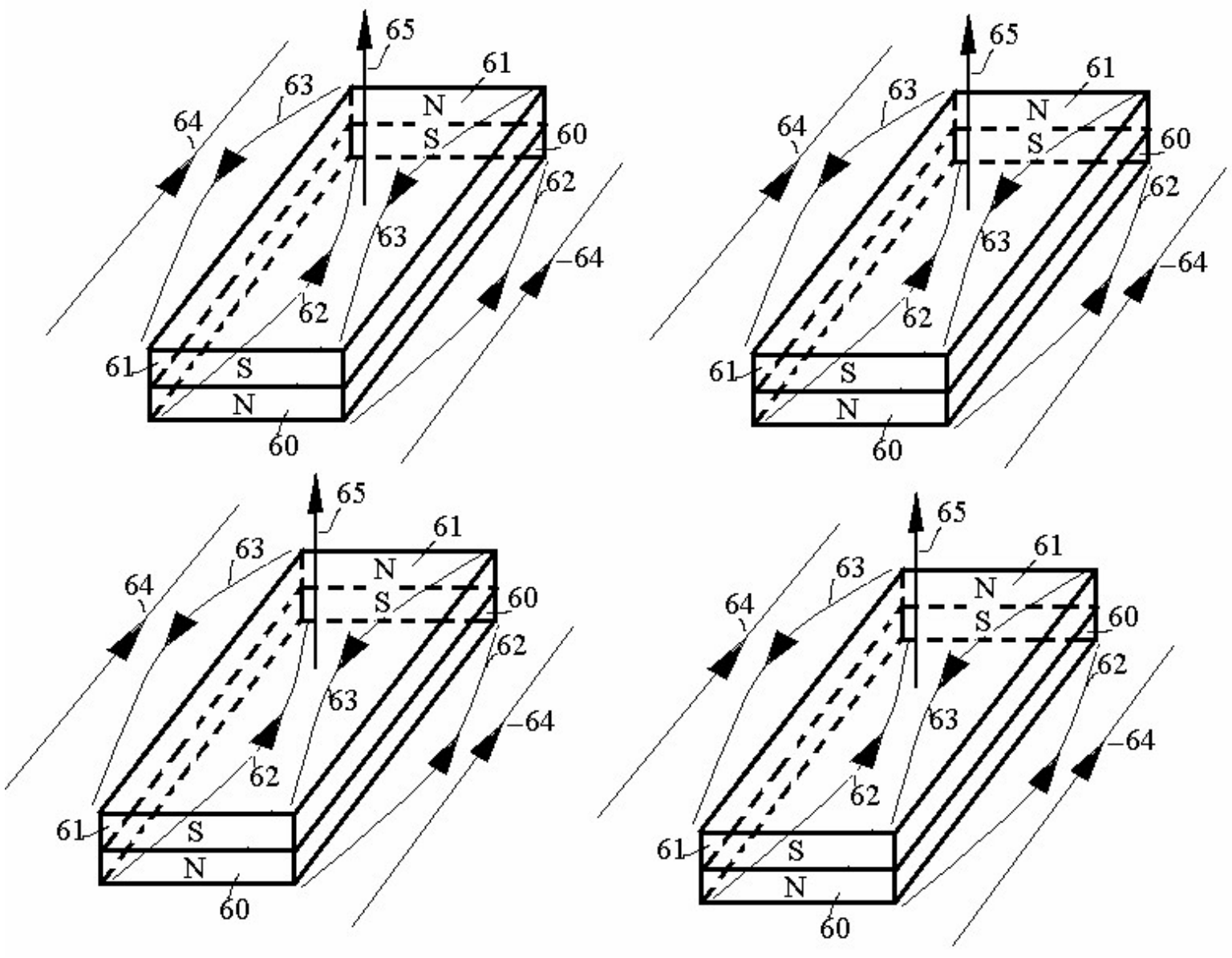

**Fig.7.**

Developed Surface Virtual Wire System. The film-like oblong magnets are arranged layer by layers on a magnetically neutral film-like base (not shown here). The sketch shows two layers. *Notations*: 60 - film-like magnets; 61 - film-like magnets having magnetic moments, which are anti-collinear to the magnetic moments of the magnets 60; 62 - magnetic fields of the magnets 60; 63 - magnetic fields of the magnets 61; 64 - external magnetic fields; 65 - net forces acting on the assemblies of the magnets 60-61.

### 2.2.4. Asymmetric Virtual Wire System

Shown in Fig.8 is a ramification of a virtual wire system, while the shown in Fig.9, induction *B* vs. magnetizing field strength *H*, explains specific features of the system, explaining its operation. The basic ramification system comprises two separated magnetic coils 70 and 71, comprising two adjacent ferromagnetic bodies: an active body 72 and a magnetic shunt 73. The coils 70 and 71 are fed with direct current, producing magnetic field 74 between them. The ferromagnetic bodies 72 and 73 are different in that the body 72 is magnetically saturated in the field 74 of the coils 70-71, while the body 73 (the magnetic shunt) is unsaturated, Fig.7. Moreover, the active body 72 has a higher magnetic induction than the magnetic shunt 73. The system rests in the external magnetic field 75. Mechanical wholeness of the system is provided by supports 76 connected to a base 77. The base 77 also serves as a source of current for the magnetizing coils 70-71. The net force 78, produced due to energetic asymmetry of the system, acts normally to the system.

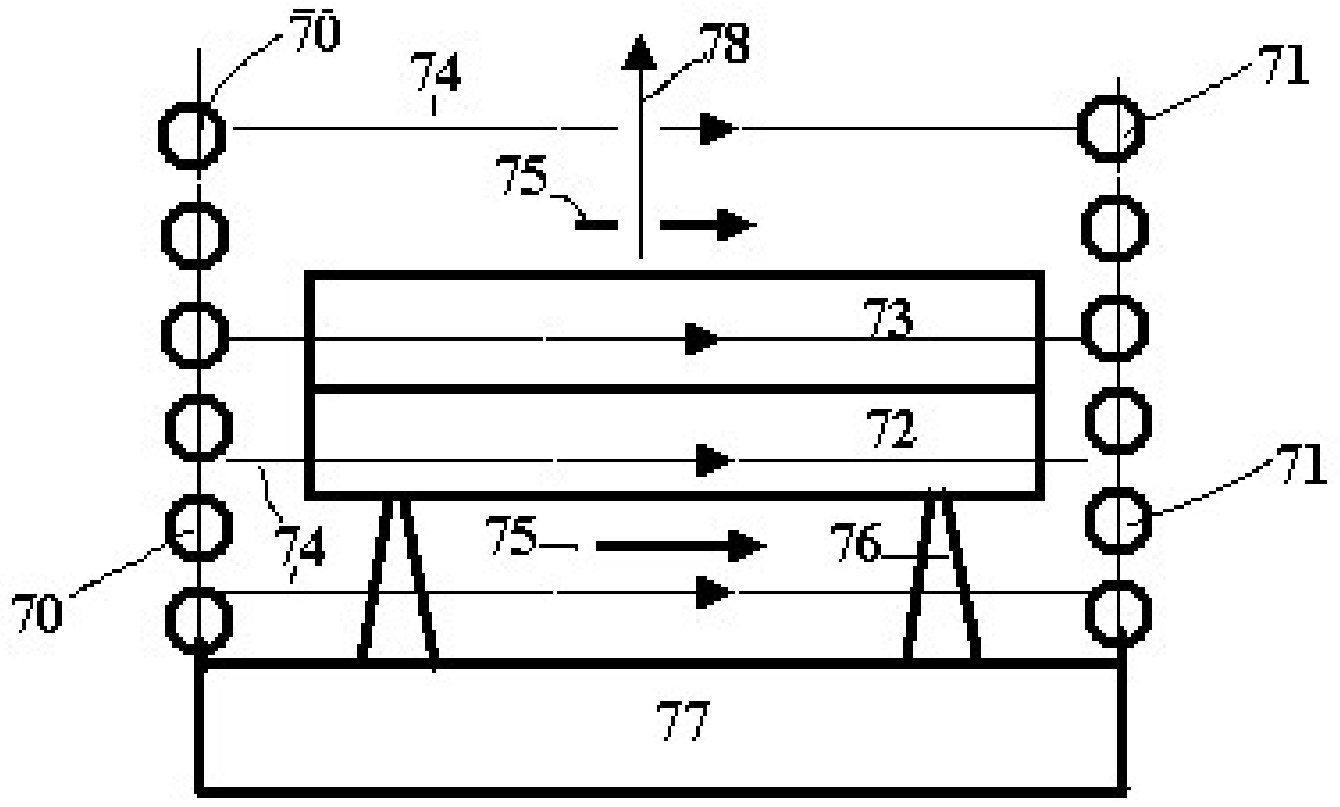

**Fig.8.**

General conception of the asymmetric virtual wire system. Notations: 70 - left magnetizing coil; 71- right magnetizing coil; 72- active body, having saturated magnetization; 73- shunt body, having unsaturated magnetization; 74- magnetic field of the coils 70-71; 75- external magnetic field; 76- mechanical supports providing wholeness of the system; 77- base comprising sources of current for the coils 70-71; 78- net force acting on the system.

The net force originates due to difference of densities of magnetic energies in immediate vicinity of the active body and the shunt. Said difference origins because of different magnetic properties of the active body and the shunt in the control field 74 of the coils 70 - 71.

**It has to be taken into consideration that only mutual magnetic energy, that is that comprising a product of external fields and that of the magnets, produces the net force**. This is a free energy of the system.

If $w_a, w_s, B_a, B_s$ and $B_e$ are densities of magnetic energies and inductions of the active body, the shunt and the induction of the external filed, then the following takes place:

$$w_a = \frac{1}{2\mu\mu_0}\left(B_a^2 + 2B_aB_e + B_e^2\right) \quad w_s = \frac{1}{2\mu\mu_0}\left(B_s^2 + 2B_sB_e + B_e^2\right),$$

$$\Delta w = w_a - w_s = \frac{1}{2\mu\mu_0}\left(B_a^2 - B_s^2 + 2B_e\left(B_a - B_s\right)\right). \tag{11}$$

Density of the free, mutual energy of the system is

$$w_f = \frac{B_e}{\mu\mu_0}\left(B_a - B_s\right) = \frac{B_e}{\mu_0\mu}\left(M_a(H) - M_s(H)\right), \tag{12}$$

where $M_a$ and $M_s$ are magnetizations, which depend on the operating field strength $H$ of the driving coils 70 - 71.

The net force is $F = \iint_S w_f dS$.

As seen from the Fig. 9, realization of this approach requires materials with characteristics like shown in this figure.

As this took place for the symmetric virtual wire system, the asymmetric one is maximally efficient as the developed surface system. Then, the elements 60s of the Fig.7 are the active bodies, while the elements 61s are magnetic shunts. Said elements made as ferromagnetic films.

*Magnetic Energy* .

Due to a vortical nature of magnetic field, $div\vec{B} = 0$, then the following is true for a flat-like bar magnet in *x-y* plane, taking into consideration variation of the field along $y$-axis

$$B_m \cdot S_m = 2\int_0^\infty \int_0^\infty B(y) dy dx \quad (13)$$

where $B_m$ and $S_m$ are induction and cross-section area of the magnetic, while *B(y)* shows the field in the external media.

Taking into consideration that for the tangential component of the field $tg\alpha_1 / tg\alpha_2 = \mu_1 / \mu_2$ and that for the **saturated magnets** $\mu \approx 1$, it's becomes obvious that a tangential inductance at a magnet-air ($\mu = 1$) border is proportional to $B_m \cdot S_m$ product.

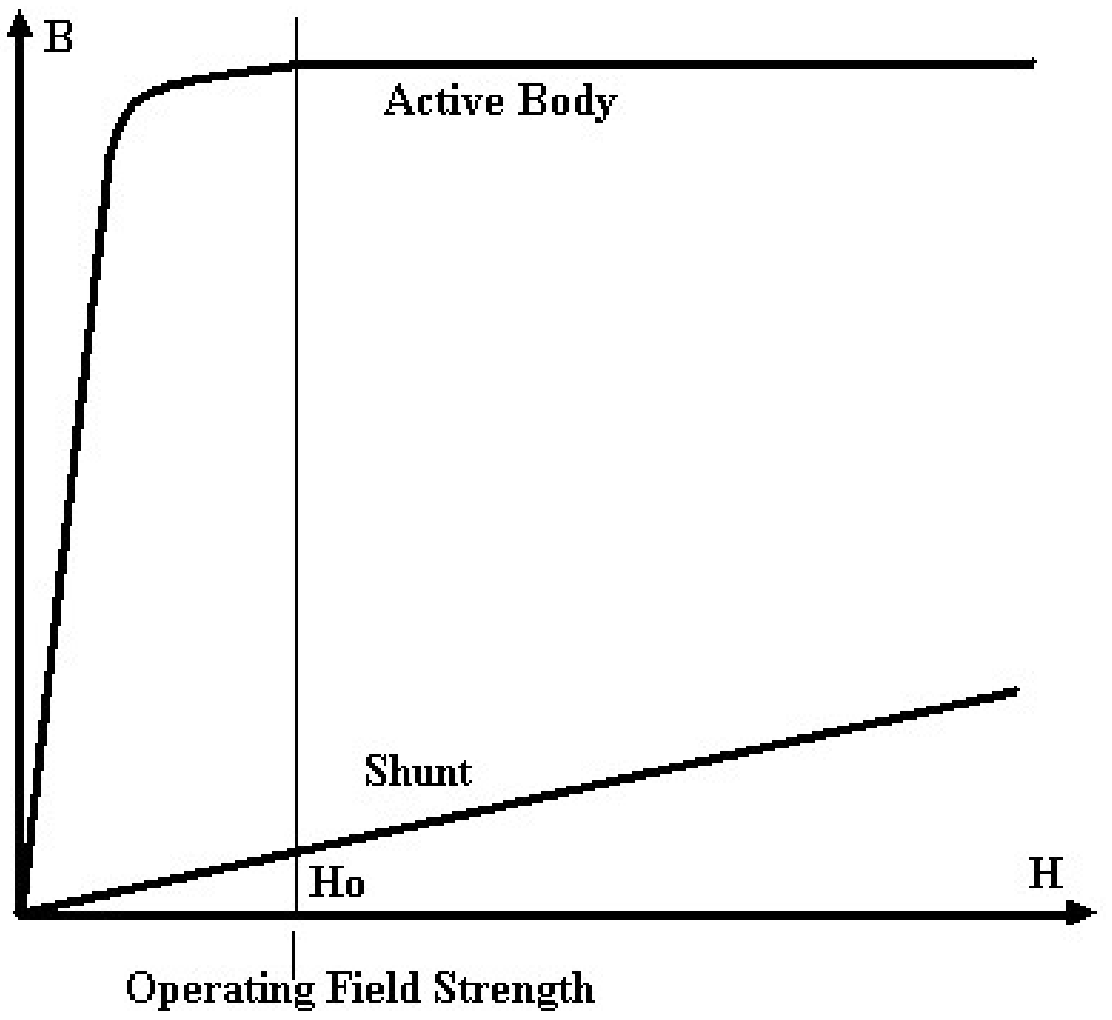

**Fig. 9.**

General magnetic properties of active body and magnetic shunt for asymmetric virtual wire system in external magnetic field.

## 2.3. Energy Loss

Moving in the external (Earth's) distance -variable magnetic field **B<sub>E</sub>**, the Virtual Wire System losses energy. As the system moves in the magnetic field, its magnets experience gradual demagnetization. The work of the displacement can be shown as

$$A = I_{eff} \cdot \Phi = I_{eff} \cdot l_{eff} \cdot \int_0^x B_E(x) dx , \qquad (14)$$

where $I_{eff}$, $l_{eff}$ are effective current and effective length of the virtual wire, while $\Phi$ is a magnetic flux, defined here as a product of Earth's magnetic induction $B_E$ and the effective area covered by the virtual conductor $l_{eff}$ while moving for a distance $X$.

The total energy loss W of the magnetic system is a sum of the lost magnetic energy and the irreversible entropy-related loss:

$$W = A + TS \qquad (15)$$

where T is a temperature, while S is an entropy of a magnetic disordering.
The magnets may be restored by a wire winding, supplied by an electric current from an internal electric source.

## 3. The Experiment

The feasibility of the proposed systems was proved experimentally. Special attention was devoted to uniformity of magnetic field.

### 3.1. The 8-Shaped Wire System

The 8-Shaped Wire System was tested in Earth's field as well as in uniform artificial field. Both mechanical and electronic balances were employed to measure the net force.

A simple experiment was made by physicist Dr. Mark Krinker showing origination of magnetic net force as it appears when a back current carrying wire is shielded in a ferromagnetic tube. The sketch of Krinker's installation is shown in Fig.10.

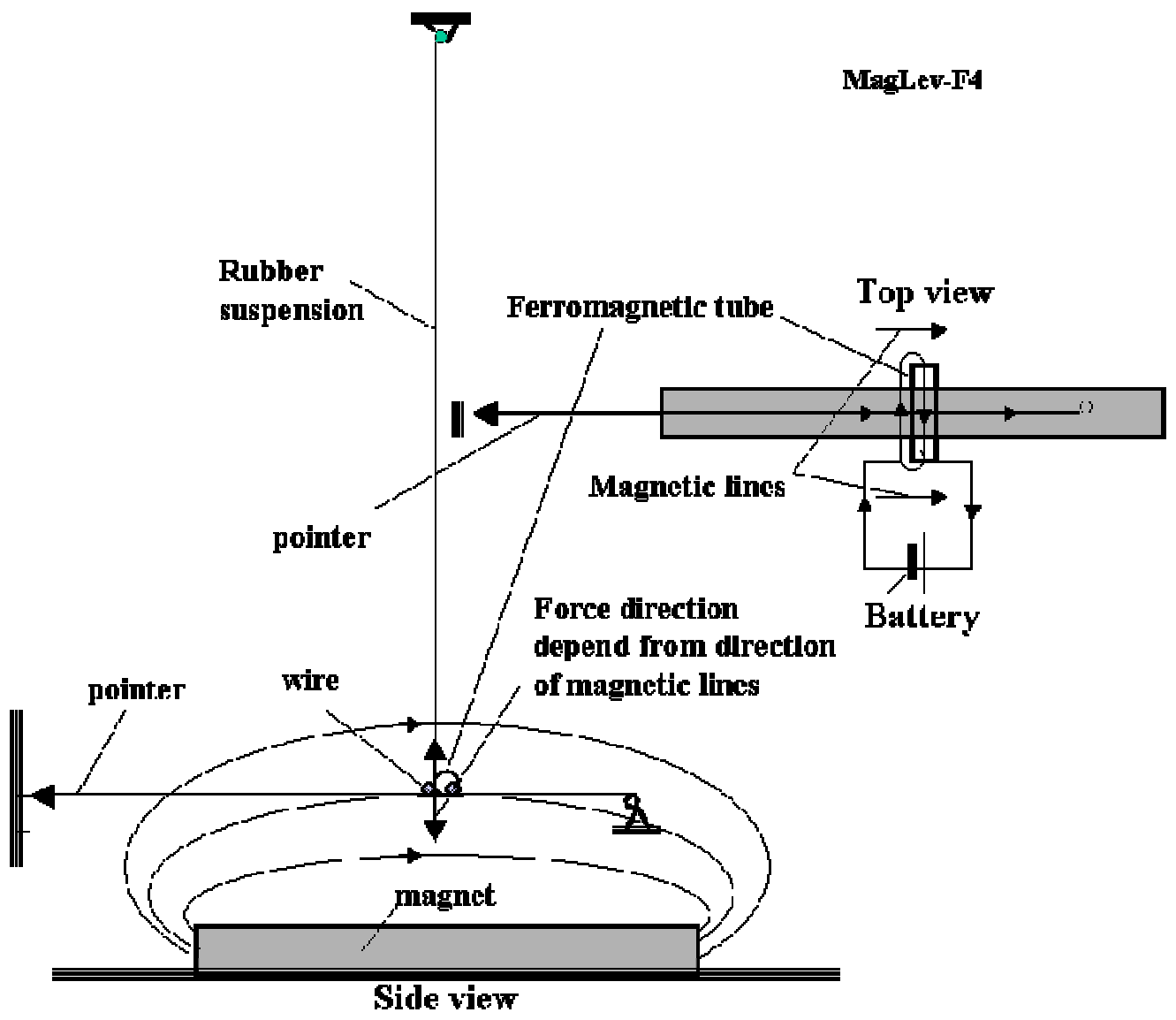

**Fig.10.** Sketch of Krinker's installation for demonstration the magnetic net force when a back current carrying wire is protected from the outer magnetic field. Data: ferromagnetic cylinder has: 28x18x10 mm; total wire of spool has length 2 m, 10 A current passed through the circuit during 7s; Magnetic field is about 0.01 T.

### 3.2. Virtual Wire System.

Special attention was devoted to uniformity of the field. For this reason, the Helmholtz coils were applied, Fig.11. However, the uniformity takes place only in central potion of the coils, as seen from the Fig.11. This is why the magnet assembly of the virtual wire system was placed on a spacer resting on the scale in a center of the coils, Fig.12. For the Virtual Wire experiment the coils with diameters 400 and 120mm respectively were employed. The results of the experiments with both the coils were video-documented.

Both the coils have 100 turns of 0.2 mm of copper wire. The big coils were supplied with DC current up to 5A and generated up to 1500μT field in the center of the installation.

The small coils, supplied with 3.6A current, generated $3.78\times10^{-3}T$ field.

The field was measured with preliminary calibrated UGN3503U linear Hall-effect sensor, Allegro Company. Beside that, the field was calculated analytically. The matching between analytical and experimental estimation was pretty good. The magnetic assembly, having total dimension 51×25×7mm, Fig.10, was made of 4 magnets. The Neodymium rare Earth magnets, having 1T internal induction were employed. The magnets were encapsulated in a plastic envelope. Two opposite directed red arrows on both sides show direction of the intrinsic magnetic fields of each the side.

**3.2.1. Active Area.**

Not all the surface of the magnetic assembly takes participation in origination of the net force: only the active area, producing horizontal lines of force in the center contributes in the effect. Those end lines of the system, which have no magnetic shunt, opposites to the effect and reduces it.

The magnetic assembly was composed of 2 pairs of thick-magnetized magnets having opposite magnetic moments (each of the pairs, in turn, was composed of two magnets). Its magnetic lines of force are shown in Fig. 4a, b. The system had no magnetic shunts.

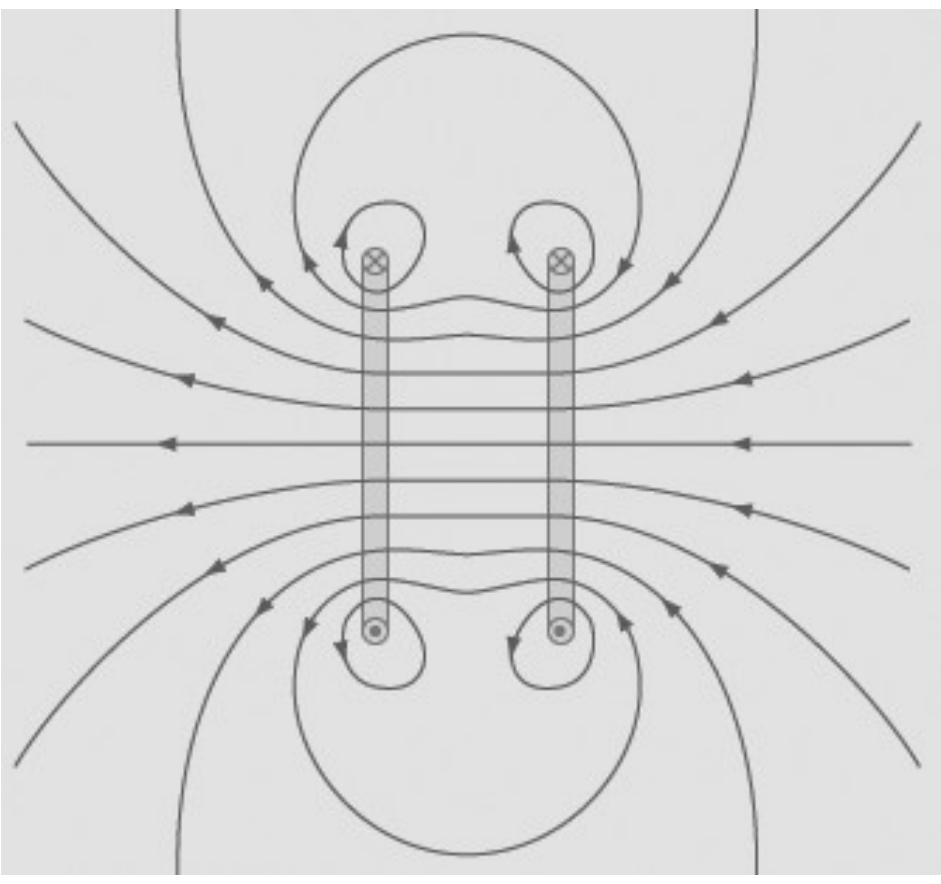

**Fig.11.** Magnetic Field Generated by Helmholtz Coils. The field is uniform in a central portion between the coils. The magnets and scales are located in the central portion.

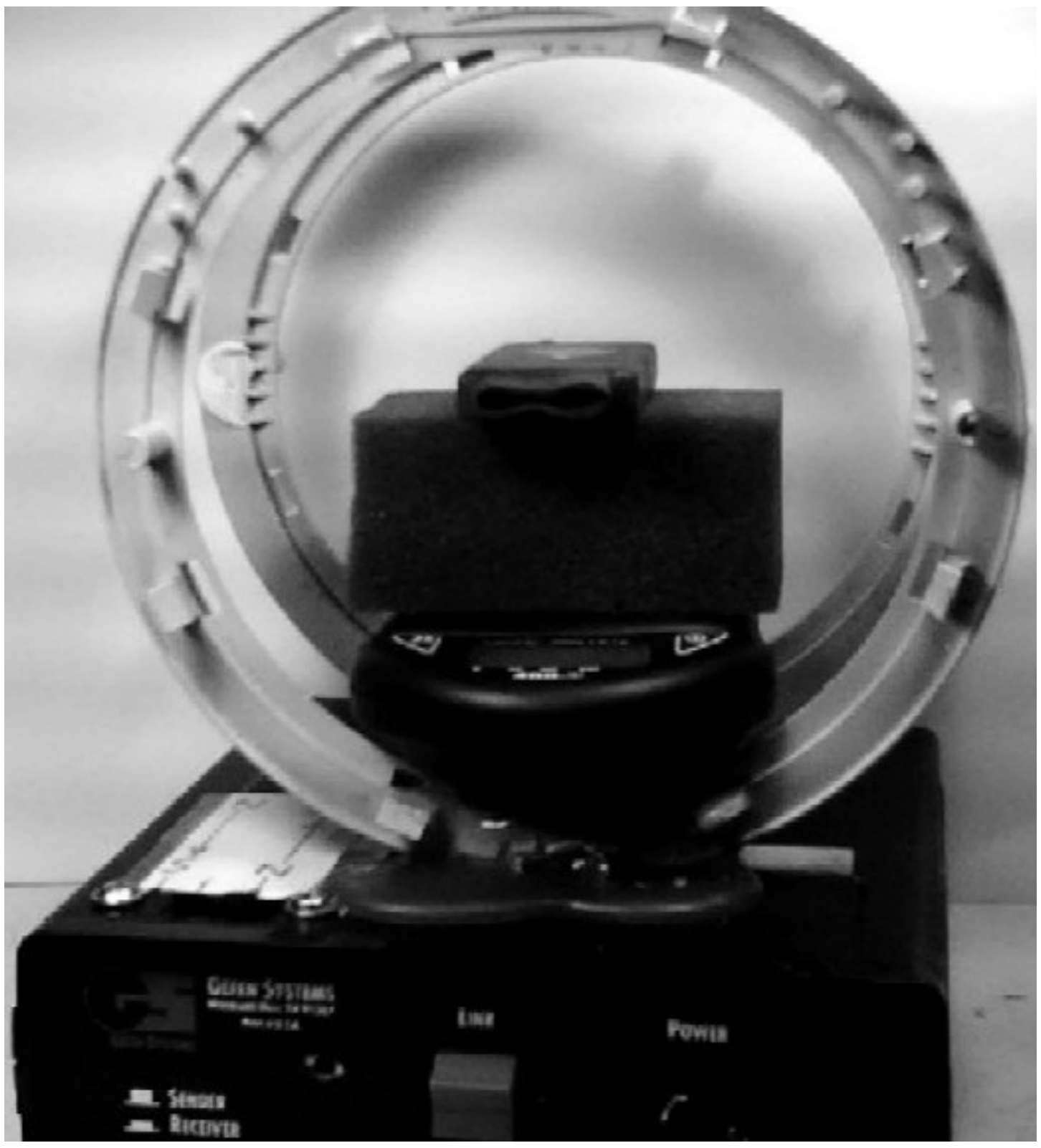

**Fig.12**.

Experimental Installation made by Dr. Krinker for the Virtual Wire System.
The Virtual Wire Magnetic Assembly (the central brick-like body on a spacer) rests on the Electronic Balance. The arrow on the magnet indicates direction of the own field of the upper side of the magnetic assembly. The direction of the field of the lower side is opposite to that of the upper side.

To study distribution of the magnetic field, iron filings in a glycerin were employed ("Doodle Pro", Fisher-Price Company). The magnetic patterns, developed by this system are shown in Fig.13a and 13b. As seen from the picture, the visible active horizontal portion (the feeling lay horizontally) counts just 5-6 mm. Because the shunts were not employed, the opposite-directed end fields existed. As shown in the Fig.13, their length equals 4 mm.

From here it follows that the length of the active area is ~2 mm. Therefore, the value of $S$ in the expression $F = \dfrac{kB_E B_M S}{\mu_0 \mu_r}$ has to be taken as a difference of the central and end portions of the magnetic patterns.

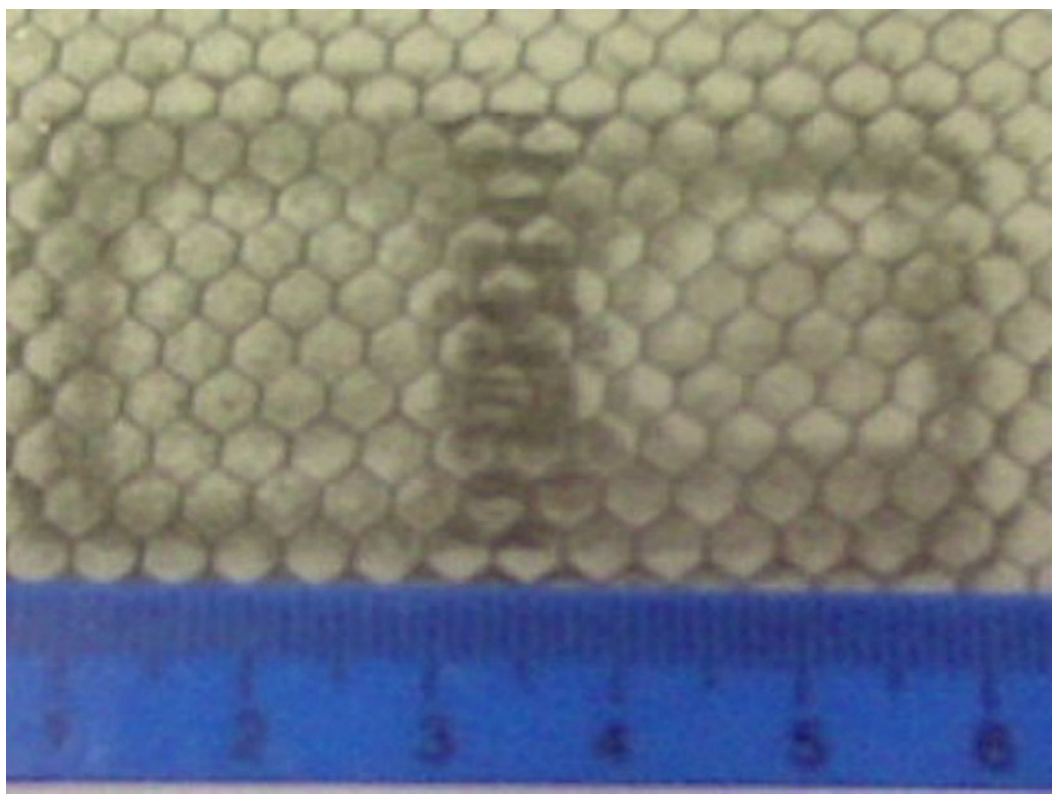

**Fig.13a**..
Patterns of intrinsic magnetic field of the magnetic assembly of Fig.13, top view, obtained with iron fillings in a glycerin.

The horizontally- oriented fillings in the center show the active area, contributing into the net force, while edge horizontal lines indicate unwanted opposite-directed field. Vertically oriented fillings (light areas) show the "idling" portion of the field.

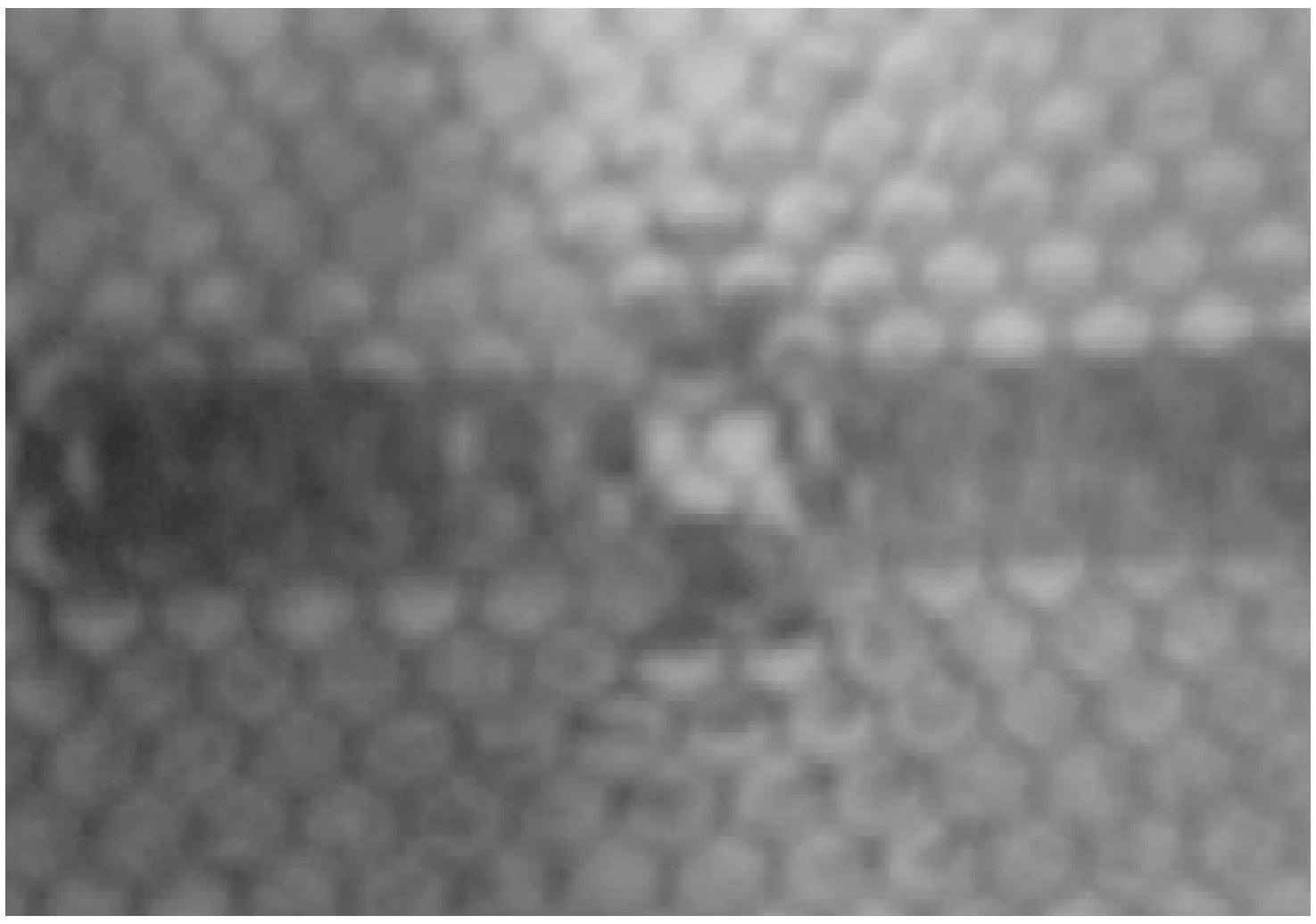

**Fig.13b.**
Side magnetic imprint of the assembly of Fig. 12. The iron filings align along closed magnetic lines of force around the center. Analogy with a powered wire is obvious.

**3.2.2 Analysis of the Experiment.**

The magnetic assembly of the Fig.12 weighs 41.7 G. Superposition field in the installation Fig.12 caused variation of the weight for 1.2 G. The sign of the variation depends on mutual orientation of the fields of the installation and the assembly.

The expected error of the experiment is $\frac{\Delta F}{F} = 2\frac{\Delta B}{B} + \frac{\Delta S}{S}$, as it follows from equation (6).

The parameters of the experiment were as follows:

The active surface: As it was shown above, after subtraction of the length of the end fields, the resulting length of the active area is 2 mm. So, the active area is

$S = 25 \cdot 10^{-3} m \times 2 \cdot 10^{-3} m = 5 \cdot 10^{-5} m^2$.

The inductions: the installation - $3.78 \times 10^{-3} T$. The measured tangential component of the induction at the surface of the magnets – 0.11T;

The following error was estimated for the experiment:

$\frac{\Delta B}{B} = 0.2$ This is a general error for measuring induction in this experiment. This error includes inaccuracy of the performed calibration as well as the error of the measuring instruments (sensor, ammeter, mili-voltmeter, caliper and so on) $\frac{\Delta S}{S} = 0.15$. As seen from the Fig.13a, measuring the active area can bring a considerable error because of illegible contours of Fig.13a. This value was obtained due to analysis of the magnetic patterns picture.

Therefore, the error for calculating net force is $\frac{\Delta F}{F} = 0.55$.

The theoretical calculation of the net force according to (6) at $k = 1$ returns

$$F = \frac{3.78 \cdot 10^{-3} T \times 1.1 \cdot 10^{-1} T \times 5 \cdot 10^{-5} m^2}{1.26 \cdot 10^{-6} H \cdot m^{-1}} = 1.65 \cdot 10^{-2} N \approx 1.68 \pm 0.93 G$$

(Here $\mu_0 = 4\pi \cdot 10^{-7} H/m \approx 1.26 \cdot 10^{-6} H/m$, while the value of the relative permeability for air $\mu_r \approx 1$)

As it follows from the data, the experimental result matches the calculation within the error of the experiment.

The divergence of the experimental result and the theoretically predictable one can be explained by inaccurate experimental values of the inductions, the active area and the geometric coefficient *k*.

**3.3. Magnetic Shunt.**

The opposite-directed side fields occupy considerable area of the magnetic assembly. Elimination of the side fields promises increasing the efficiency of the generated net force. Figs. 14-15 show effect of the magnetic shunting of the side fields. The simple magnetic shunt is made of ferromagnetic tubes attached to the end of the thick-magnetized magnet.

. The shunt“absorbs” magnetic lines of force, Fig.15 compared to Fig 14.

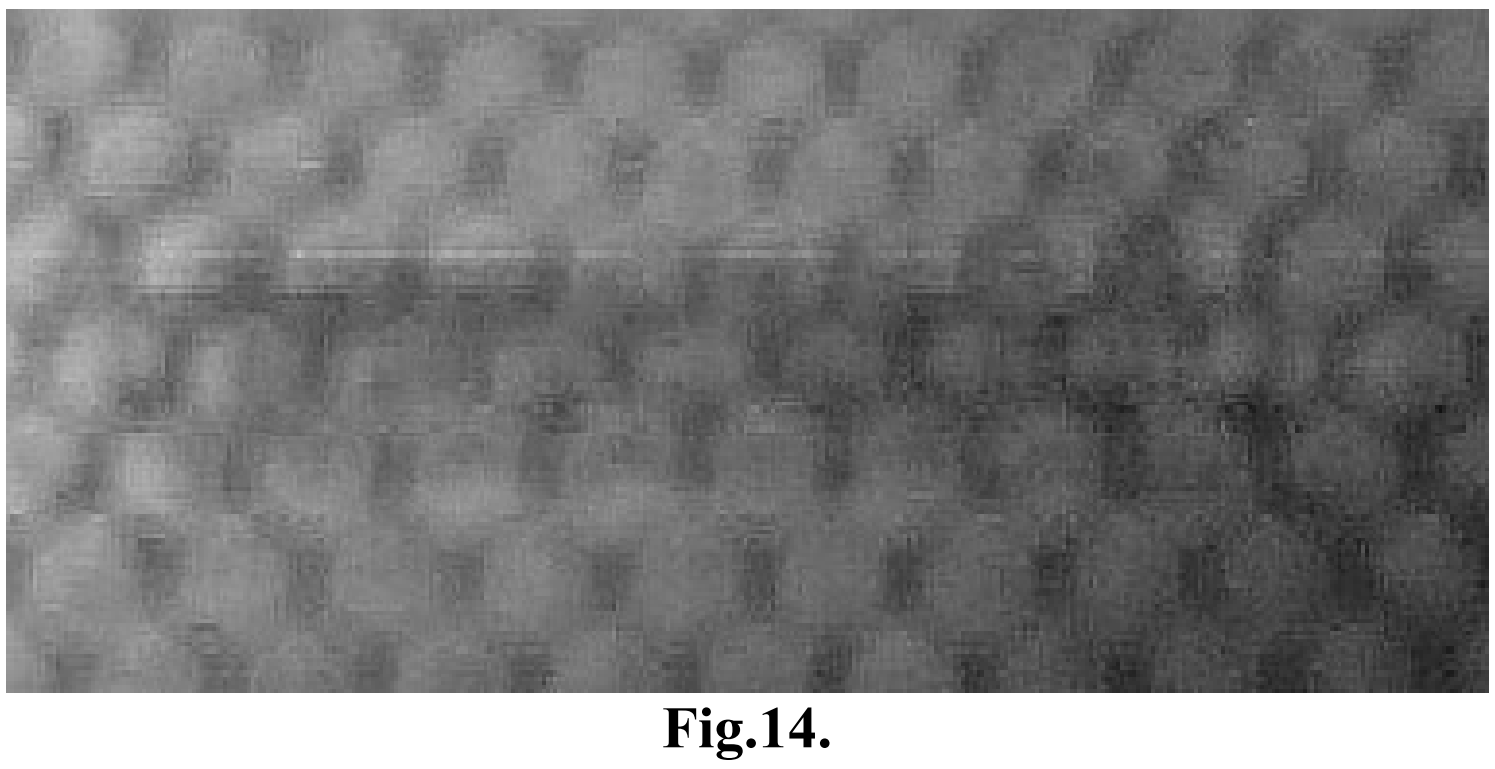

**Fig.14.**

An imprint of the side magnetic field of the magnetic assembly of Fig.12
(the dark strip in the center)

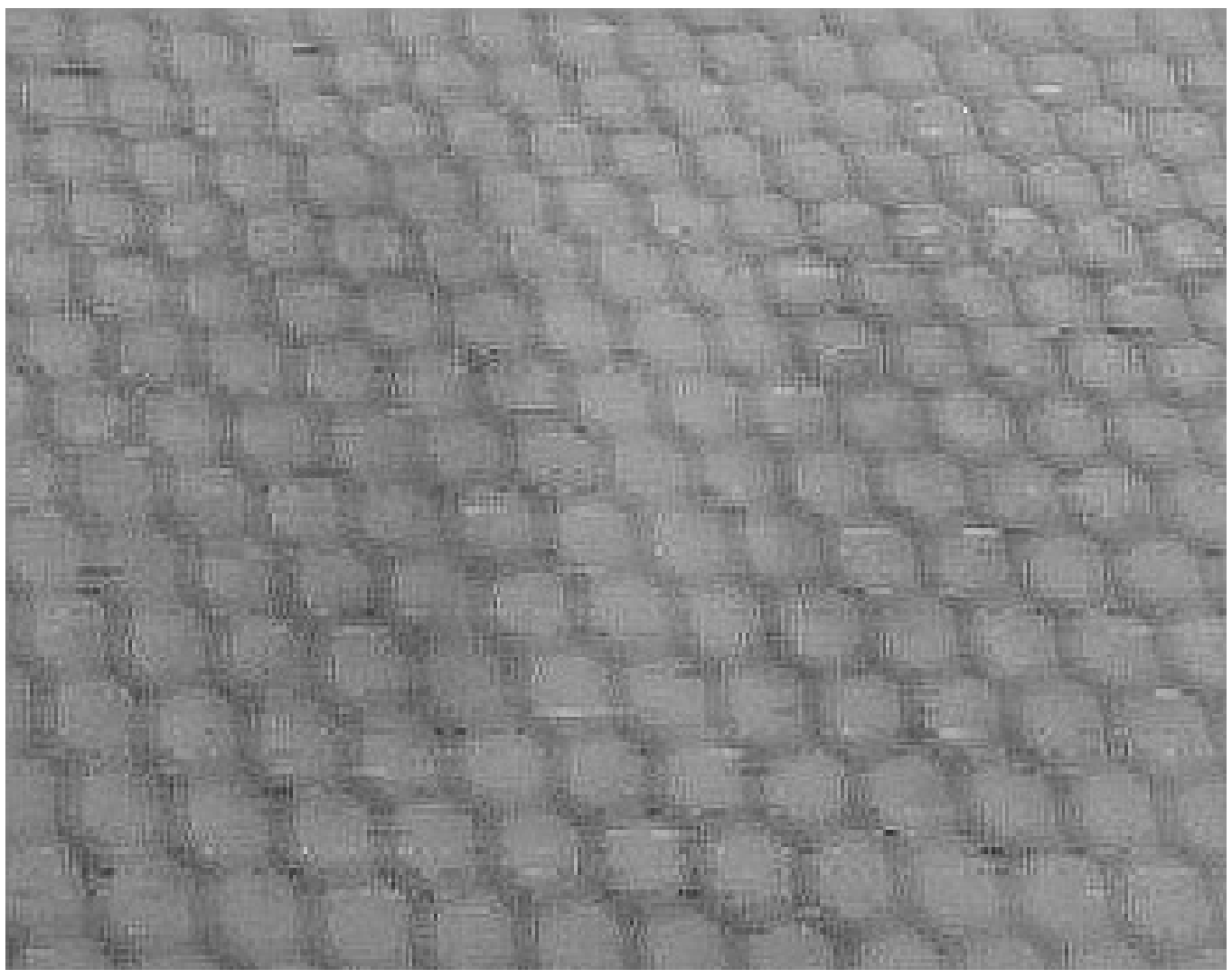

**Fig.15.**
Influence of the magnetic shunt: a ferromagnetic tube, attached to the side, guides the magnetic lines of force, living no imprint in the iron filings (the light strip in the center)

## Reference

1. G. Wouch and A.E. Lord, Jr. Levitation, Electromagnetic, Encyclopedia of Physics, p. 627, VCH Publishers, INC, Second Edition.
2. Krinker, M., Magnetofly-1, Disclosure Document No. 269248, US Patent &Trademark Office, 1990.
3. Krinker, M., Permanent Magnets Based Maglev, Disclosure Document No. 516184, US Patent & Trademark Office, 2002.
4. Bolonkin, A.A., "Theory of Flight Vehicles with Control Radial Force". Collection *Researches of Flight Dynamics.* Mashinostroenie Publisher, Moscow, 1965, pp.79-118 (in Russian). International Aerospace Abstract A66-23338# (in English).
   The work contains theory of flight vehicles with control gravity or radial Magnetic force.
5. Bolonkin, A.A., "Electrostatic Levitation on the Earth and Artificial Gravity for Space Ships and Asteroids". Paper AIAA-2005-4465, 41-st Propulsion Conference, 10-13 July 2005, Tucson, AZ, USA.
   The work contains theory of electroctatic levitation of flight vehicles and creating artificial gravity in space ships without rotation of apparatus (see also [6], Ch. 15).
6. Bolonkin, A.A., Non-Rocket Space Launch and Flight, Elsevier, 2006, 488 pgs.
   The book contains theories of the more then 20 new revolutionary author ideas in space and technology.
7. Bolonkin, A.A., AB Levitator and Electricity Storage, This work presented as paper AIAA-2007-4612 to 38th AIAA Plasmadynamics and Lasers Conference in conjunction with the16th International Conference on MHD Energy Conversion on 25-27 June 2007, Miami, USA. See also http://arxiv.org search "Bolonkin".
8. Bolonkin, A.A., AB Levitrons and their Applications to Earth's Motionless Satellites. See: http://arxiv.org search "Bolonkin", July, 2007.
9. Bolonkin, A.A., New Concepts, Ideas, Innovation in Aerospace, Technology and Human Science, NOVA, 2008, 320 pgs. Ch.1, Part 1. AB Levitation and Electricity Storage. Ch. 12, Part 2. AB Levitation and their Application.
10. Bolonkin, A.A., AB Electron Tube and Semi-Superconductivity at Room Temperature. See http://arxiv.org search "Bolonkin".
11. *Technology to Youth Magazine,* Reports of Inversor Laboratory, Moscow, April, 1971.
12. Krinker, M., To Possibility of Human Levitation. Published in Bio-Field Interactions and Medical Technologies. International Science Conference, pp.29-33, Moscow, April 2008.
13. Bolonkin A.A., and Cathcart R.B., Macro-Projects: Technology and Environment, NOVA, 2008, 480 pgs.